\documentclass[%
aip,%
amsmath,amssymb,floatfix,
reprint,%
jcp,%
longbibliography,
]{revtex4-1}

\usepackage{graphicx}
\usepackage{dcolumn}
\usepackage{bm}
\usepackage{amssymb}
\usepackage{mathtools}
\usepackage{multirow}
\usepackage{bbold}
 \usepackage[table,xcdraw]{xcolor}
\usepackage{hhline}

\newcommand{\1}{\begin{equation}}
\newcommand{\2}{\end{equation}}
\newcommand{\ea}{\begin{eqnarray}}
\newcommand{\ee}{\end{eqnarray}}

\begin{document}


\title{Inertial effects of self-propelled particles: from active Brownian to active Langevin motion}



 \author{Hartmut L\"owen}
 \affiliation{Institut f\"{u}r Theoretische Physik II: Weiche Materie, Heinrich-Heine-Universit\"{a}t D\"{u}sseldorf, D-40225 D\"{u}sseldorf, Germany}


\date{\today}

\begin{abstract}
Active particles which are self-propelled by converting energy into mechanical motion represent an
expanding research realm in physics and chemistry. For micron-sized particles moving in a liquid (``microswimmers''),
most of the basic features have been described by using the model of overdamped active Brownian motion.
However, for macroscopic particles or microparticles moving in a gas, inertial effects become relevant
such that the dynamics is underdamped. Therefore, recently, active particles with inertia have been
described by extending the active Brownian motion model to active Langevin dynamics which include inertia.
In this perspective article recent developments of active particles with inertia (``microflyers'') are summarized
both for single particle properties and for collective effects of many particles. There include: inertial delay
effects between particle velocity and self-propulsion direction, tuning of the long-time self-diffusion by
the moment of inertia, effects of fictitious forces in non-inertial frames, and the influence of inertia on
motility-induced phase separation. Possible future developments and perspectives are also proposed and discussed.
\end{abstract}


\maketitle 

Invited \textbf{perspective article} for \textit{The Journal of Chemical Physics}.

\section{Introduction}\label{ra_sec1}

The dynamics of self-propelled particles which are perpetously moving by converting energy ("fuel") into mechanical motion
represent a nonequilibrium phenomenon. Research in the last decades was not only driven by the broad range of applications
(such as precision surgery, drug delivery and cargo transport on the micron-scale) but also from a more fundamental level
in terms of identifying basic relevant models to describe the particle trajectories under nonequilibrium conditions.
One of the first standard models in this respect is the Vicsek model of swarming proposed in 1995 by Vicsek and coworkers \cite{Vicsek}
which is by now a cornerstone in describing collective active matter systems. With the upsurge of synthetic colloidal Janus-like
particles which create their own gradient in which they are moving, artificial microswimmers were considered as model systems
for active matter. These micron-sized particles typically self-propel in a liquid at very low Reynolds number. Therefore the
dynamics of these colloidal particles in a solvent is overdamped and one of the most popular descriptions is obtained by
{\it active Brownian motion\/} \cite{GolestanianPRL2007,Hagen2009,tenHagenJPCM} combining
solvent kicks described as Gaussian white noise and overdamped motion together with an effective self-propulsion force
representing the particle self-propulsion. The  active Brownian particle  model has been tested against experimental
data of self-propelled colloids \cite{GolestanianPRL2007,Kuemmel_et_al_PRL_2013,Kurzthaler_Poon}
and has been used also to describe and predict collective phenomena for colloids and bacteria \cite{ourRMP}.

More recently, there have been developments to consider larger self-propelled particles or motion in low-density environment
(gas instead of liquid). Then the motion is not any longer at low Reynolds number. Instead inertial effects are getting relevant
in the dynamics and need to be included in the modelling. These inertia-dominated particles are "microflyers" rather than
"microswimmers" since their dynamics is underdamped and rather corresponds to flying than swimming.
Still the motion is affected by fluctuating random kicks of the surrounding medium.
Correspondingly
one can coin their dynamics as {\it active Langevin motion\/} rather than "active Brownian motion". However, it is remarked here that
sometimes
in the literature \cite{Ebeling,Romanchuk_review} the term "active Brownian motion" is used in a more general sense
including also  underdamped Langevin equations of motion.

Examples are mesoscopic dust particles in plasmas (so-called "complex plasma") \cite{Ivlev_Morfill}.
Pairs of such dust particles can be brought into a joint self-propulsion \cite{Bartnick2} by non-reciprocal interactions
induced by ionic wake charges \cite{Bartnick1}. These motions are only virtually damped. Another important realization are
granulates made self-propelling on a vibrating plate \cite{Ramaswamy,Dauchot_Frey_PRL_2013,Poeschel,Deblais,Tsimring,Chate,Patterson,Dauchot2,Mayya1,Mayya2} or equipped with an internal vibration motor \cite{Dauchot_Demery_PRL_2019} where it has been shown that
the active Langevin model indeed describes their dynamics well \cite{Walsh,Dauchot_Brownian,Ldov}.
Further examples for self-propelled particles with inertia range from  mini-robots \cite{Mijalkov,Leyman} to
macroscopic swimmers like beetles flying at water interfaces \cite{beetles=Ref[24]_in_Mandal}
and whirling fruits self-propelling in air \cite{beetles=Ref[25]_in_Mandal}.

In this perspective article, we first briefly review basic features and predictions of active Brownian motion
discussing both swimmers moving on average on a straight line ("linear swimmers") and particles swimming on a  circle
("circle swimmers"). Both single particle properties and collective effects such as
motility-induced phase separation are briefly discussed. We then extend the model towards active Langevin motion
including inertia and summarize recent developments. We show that some of the basic properties of active Brownian
motion are qualitatively changed due to inertia. In particular we discuss inertial delay effects between particle
velocity and self-propulsion direction, the tuning of the long-time self-diffusion by the moment of inertia,
effect of fictitious forces in non-inertial frames, and the influence of inertia on motility-induced phase
separation (MIPS). MIPS is strongly influenced and suppressed by inertia and if there are two coexisting phases
of high and low particle density, these coexisting phases possess different kinetic temperatures.

The paper is organized as follows: in section II, we first recapitulate the basic features of active
Brownian motion both for linear swimmers and for circle swimmers. Then, in section III, we propose and
discuss the model of active Langevin motion

\section{Active Brownian motion for self-propelled colloidal particles (``microswimmers'')}\label{ra_sec2}

Let us first recapitulate the basic features for overdamped active Brownian motion. In the $xy$-plane,
a single particle trajectory at time $t$ is described by the particle center ${\vec r}(t) = (x(t), y(t))$ and
particle orientation $\hat n(t) = (\cos \phi (t) , \sin \phi (t))$  where $\phi( t)$ is the angle of the particle orientation
with the $x$-axis. We now distinguish between linear swimmers and circle swimmers which
experience a systematic torque and exhibit chirality.

\subsection{Linear swimmers}\label{CIIa}

For linear swimmers, the basic overdamped equations of motion read as
\begin{equation} \label{eq:1}
\gamma \dot{\vec{r}}(t) = \gamma v_0 \hat{n}(t) + \vec{f}(t)
\end{equation}
\begin{equation} \label{eq:2}
\gamma_{R}{\dot \varphi} = g(t)
\end{equation}
These equations couple translational and rotational motion and represent a force and torque balance.
In detail,  $\gamma$ denotes a translatorial friction coefficient,
$v_0$ is the imposed self-propulsion speed of the active particle and $\gamma_R$ is a rotational friction coefficient.
The components of ${\vec f} (t)$ and $g(t)$ are Gaussian random numbers
with zero mean and variances representing white noise from the surrounding, i.e. $\overline{\vec{f}(t)}=0$, $\overline{f_i(t)f_j(t')}=2 k_B T \gamma \delta_{ij}\delta(t-t')$, $\overline{g(t)}=0$, $\overline{g(t)g(t')}=2 k_B T_R \gamma_R \delta(t-t')$ where the overbar means a noise average.
Here $k_BT$ denotes an effective thermal energy quantifying the translational noise strength. Likewise
$k_BT_R$ characterizes the orientational noise strength. In many applications, these temperatures are set to be equal, i.e.\
$T\equiv T_R$, in others the translational noise is neglected with respect to the rotational noise such that $T=0$ and $T_R>0$.
In the noise-free case $T=T_R=0$, the self-propelled motion is linear along the particle orientation,
i.e. the self-propelled particle is a
{\it linear swimmer\/}. Its propulsion speed is the imposed $v_0$.
As a remark, the orientational dynamics can equivalently be written as
\begin{equation} \label{eq:equivalent}
\gamma_R \dot {\hat{n}}  =  {\vec g}(t)  \times {\hat{n}}
\end{equation}
where we extended all vectors to three dimensions such that $\hat{n} = (n_x,n_y,0)$ and
${ \vec g}(t) = (0,0,g(t))$.

Let us first discuss the number of independent parameters inherent in the equations of motion (\ref{eq:1}) and (\ref{eq:2}) in the steady state. By  choosing suitable units for length and time
such as the persistent length $\ell_p = v_0/D_r$ and the persistence time  $\tau_p=1/D_r$
where  $D_r=k_BT_R/\gamma_R$ denotes the rotational diffusion constant, a scaling of the equations results in only one
remaining independent parameter
which can be chosen to be a dimensionless {\it Peclet number}
\begin{equation} \label{eq:Peclet}
Pe = \gamma v_0 \sqrt{\gamma \gamma_R/k_B^2TT_R}
\end{equation}
This Peclet number measures the strength of activity with respect to noise, it
diverges for the zero translational noise
such that in this case the model  set by Eqns. (\ref{eq:1}) and (\ref{eq:2})
is a completely  parameter-free persistent random walk solely characterized by the
persistence length and persistence time.

We now address the noise-averaged displacement as a function of time $t$ for a prescribed
initial orientation ${\hat n}(0)$ at time $t=0$. It is given by \cite{Houwse,tenHagenJPCM}
\begin{equation} \label{eq:18}
\overline{\vec{r}(t)-\vec{r}(0)} = \frac{v_0}{D_r}(1-e^{-D_r t})\hat{n}(0)=\ell_p(1-e^{-t/\tau_p}) {\hat n}(0)
\end{equation}
This represents a linear segment oriented
along ${\hat n}(0)$ whose total length is the persistence length. The intuitive interpretation of Eq. (\ref{eq:18})
is that due to the coupling between translational and rotational motion the trajectories show a persistence, it is a persistent random walk rather than a standard random walk. The particle remembers where it came from since it is self-propelled along its orientation and the orientation diffuses with $D_r$. It is the orientational fluctuations which are governing the persistence not the translational fluctuations. The motion is therefore a ``random drive'' rather than a ``random walk'': A blind driver steers a car with fluctuations in the steering wheel orientation, and this is what makes the motion persistent.

Next we can calculate the mean-square displacement (MSD) which is analytically given by
\cite{Houwse,tenHagenJPCM}:
\begin{equation} \label{eq:20a}
\begin{aligned}
\overline{(\vec{r}(t)-\vec{r}(0))^2} &= \frac{v_0^2}{D_r^2}(D_rt-1+e^{-D_rt})
+ 4Dt
\end{aligned}
\end{equation}
Of course it does not depend on the initial orientation ${\hat n}(0)$ due to symmetry.
Expanding Eq.\ (\ref{eq:20a}) for small, intermediate and long times, we obtain the diffusive short-time
limit for the MSD as $4Dt$ where $D$ is the translational short-time diffusion
which can be expressed as $D=\ell_p^2 / \tau_p Pe$. The initial diffusive regime
is then followed by a ballistic regime at intermediate times where the
MSD scales roughly as $(v_0 t)^2$. For long times, the MSD is again diffusive but with a much larger long-time diffusion coefficient
\begin{equation} \label{eq:20b}
D_L = \lim_{t \to \infty} \frac{1}{4t} \overline{(\vec{r}(t)-\vec{r}(0)}^2
= D + v_0^2/4D_r = ( 1/Pe +\frac{1}{4})\ell_p^2 / \tau_p 
\end{equation}
Remarkably, for strong self-propulsion, $D_L$ is much larger than $D$ (i.e. $Pe \to \infty$)
such that $D_L=\ell_p^2 / \tau_p$ consistent with the persistent random walk picture.

It is important to note that in the overdamped Brownian model the velocities $\dot{\vec{r}}(t)$ are
not real observables as they fluctuate without any bounds due to the noise terms in Eq.\ (\ref{eq:1}).
Instead can define an averaged or drift velocity by
${\vec v}_d (t) = \lim_{\Delta t \to 0} \overline{(\vec{r}(t+\Delta t)-\vec{r}(t))}/\Delta t$
which is given ${\vec v}_d (t)=v_0 {\hat n}(t)$, i.e. the systematic part of the particle drift velocity
is the self-propulsion velocity. However, still one can define a velocity autocorrelation function $Z(t)$ as a
second time-derivative of the mean-square displacement \cite{Hansen_Mac_Donald_book_Theory_of_Simple_Liquids} 
\begin{equation} \label{eq:Z}
Z(t) =\frac{d^2}{dt^2} \overline{(\vec{r}(t)-\vec{r}(0))^2}
\end{equation}
In the case of active Brownian motion Eq.\ (\ref{eq:20a}) yields that the velocity autocorrelation function is a
\textit{single exponential}
\begin{equation} \label{eq:Z1}
Z(t) =v_0^2 \exp( - D_r t)
\end{equation}
decaying with the persistence time $\tau_p=1/D_r$. This result also implies that the mean squared velocity
is the self-propulsion speed as in the short-time limit $ Z(0)= v_0^2$.

Remarkably the orientational correlation function $C(t)$ is also a {\it single exponential\/}
\begin{equation} \label{eq:20c}
C(t) = \overline{\hat{n}(t)\cdot\hat{n}(0)} = \exp( - D_r t)
\end{equation}
decaying with the same persistence time $\tau_p=1/D_r$, i.e. it is proportional to $Z(t)$. This
documents that we have standard Brownian orientational diffusion in two dimensions \cite{JanDhontbook_Dynamics_of_colloids}.


Finally one finds for the correlation function between particle orientation and drift velocity the result
\begin{equation} \label{eq:20delay}
\begin{aligned}
c(t',t) &= \lim_{\Delta t \to 0} \overline{\hat{n}(t')\cdot (\vec{r}(t+\Delta t)-\vec{r}(t))/\Delta t } \\
&= v_0\exp( - D_r |t-t'|)
\end{aligned}
\end{equation}
and a delay function which measures how the dynamical changes of orientation and velocities are correlated can be defined
via
\begin{equation} \label{eq:20delay2}
d(t) = c(t,0) - c(0,t) \equiv 0
\end{equation}
The delay function trivially vanishes here by symmetry  but this will not hold for inertia as discussed later.


Experimental data for the noise averages obtained for dilute self-propelled colloidal
Janus particles could indeed by described with these predictions
\cite{Houwse,Silber_Li} establishing the active Brownian motion model for active colloidal particles.

\subsection{Circle swimmers}\label{CIIb}

In practice, microswimmers are not perfectly rotational symmetric around their swimming axis. An asymmetry leads to a
systematic circular motion or chiral motion. In two spatial dimensions, this has been described by including an
effective  torque $M$ in the equations of motion
\cite{van_Teeffelen_Loewen_PRE_2008} as
\begin{equation} \label{eq:1cs}
\gamma \dot{\vec{r}}(t) = \gamma v_0 \hat{n}(t) + \vec{f}(t)
\end{equation}
\begin{equation} \label{eq:2cs}
\gamma_{R}{\dot \varphi} = M + g(t)
\end{equation}
The noise-free trajectories are circles with  a spinning frequency
\begin{equation} \label{eq:12}
\omega_s = \frac{M}{\gamma_R}
\end{equation}
and a spinning radius
\begin{equation} \label{eq:13}
R_s = \frac{ v_0 \gamma_R}{M}
\end{equation}
The sign of the torque $M$ determines whether the circling motion is clockwise or anti-clockwise.
Now compared to Eqns.\ (\ref{eq:1}) and (\ref{eq:2})  there is an additional independent system
parameter which can be chosen as an additional
reduced time scale related to the spinning frequency $ \omega_s\tau_p =\frac{M}{\gamma_R}\tau_p $.

Again the mean displacement can be calculated analytically and turns out to be
a logarithmic spiral ({\it "spira mirabilis"}) given by \cite{van_Teeffelen_Loewen_PRE_2008}:
\begin{equation} \label{eq:19}
\begin{aligned}
\overline{\vec{r}(t)-\vec{r}(0)} &= \lambda (D_r \hat n(0) + \omega_s \hat n^{\perp}(0)  \\
&- e^{-D_r t} (D_r \bar{\hat n} + \omega_s \bar{\hat n}^{\perp}))
\end{aligned}
\end{equation}
with $\lambda = v_0/(D_r^2 + \omega_s^2)$
\begin{equation*}
\begin{aligned}
\hat n^{\perp}(0) &= (-\sin{\phi (0)}, \cos{\phi (0)}) \\
\bar{\hat n} &=(\cos{(\omega_s t+ \phi(0))}, \sin{(\omega_s t +\phi(0))}), \;\text{and}\\
\bar{\hat n}^{\perp}& =(-\sin{(\omega_s t + \phi(0))}, \cos{(\omega_s t + \phi(0))})
\end{aligned}
\end{equation*}

Likewise the MSD for circle swimmers  is obtained as
\cite{van_Teeffelen_Loewen_PRE_2008}:
\begin{equation} \label{eq:20}
\begin{aligned}
\overline{(\vec{r}'(t)-\vec{r}'(0))^2} &= 2 \lambda^2  \left[\omega_s^2 - D_r^2 + D_r (D_r^2+\omega_s^2)t+ e^{-D_r t}\right. \\
 & \left.\cdot [(D_r^2-\omega_s^2)\cos{\omega_s t}-2D_r \omega_s \sin{\omega_s t}]\right] + 4Dt
\end{aligned}
\end{equation}
which is diffusive for both short times and long times with the short-time diffusion coefficient
$D$ and the long-time diffusion coefficient
\begin{equation} \label{eq:20long}
D_L=D + \frac{v_0^2D_r}{4(D_r^2 + \omega_s^2)}
\end{equation}
which implies that circular spinning will reduce the long-time diffusion coefficient.

From  Eq.\ (\ref{eq:Z}) the velocity autocorrelation function can be calculated  as
\begin{equation} \label{eq:20}
Z(t) = v_0^2 \cos(\omega_s t) \exp( - D_r t)
\end{equation}
It is not a single exponential but contains a further time scale $1/\omega_s$
reflecting the systematic spinning of the particle orientation.

Finally the orientational correlation function is
\begin{equation} \label{eq:20cs}
C(t) = \overline{\hat{n}(t)\cdot\hat{n}(0)} = \cos (\omega_s t) \exp( - D_r t)
\end{equation}
It is proportional to $Z(t)$ and contains the further time scale $1/\omega_s$ of particle circling as well.

For anisotropic colloidal self-propelled Janus particles, the active Brownian model was tested in detail.
In particular, the {\it spira mirabilis\/}  for the mean displacement was confirmed \cite{Kuemmel_et_al_PRL_2013}.

The active Brownian motion of a single particle can be generalized to  the presence of
external potentials (such as confinement or gravity) and external
flow fields (such as linear shear flow), for a review of the different situations considered, see
\cite{ourRMP}.

\subsection{Collective effects of active Brownian particles: MIPS}\label{CIIc}

Active Brownian particles exhibits a wealth of fascinating collective effects
including swarming \cite{Vicsek}, clustering \cite{Zoettl_Stark_JPCM,ourRMP}, crystallization
\cite{Speck2012,Pagonabarraga2018?,Krauth_1,Krauth_2}  and turbulence \cite{Wensink_PNAS}.
Here we focus more on one important effect that is purely induced by activity and
is called
motility-induced phase separation (MIPS). It was seen in computer simulations of active Brownian particles without aligning interactions
\cite{FilyMarchettiPRL2012,Buttinoni_PRL_2013} and confirmed in experiments on artificial colloidal Janus-particles
\cite{Buttinoni_PRL_2013,Sacanna_Pine_Science_2013}.

The basic idea behind MIPS is as follows:  Consider a system of active Brownian particle
at a finite concentration where the particle are interacting by purely repulsive non-aligning interactions.
Suppose two particles meet in a perfect central
collision as sketched in Figure 1a. They will not bounce back but stay close to eachother until rotational
diffusion will turn the orientations away such they can pass along eachother. This process will
happen on a time scale of $\tau_p=1/D_r$. If other particles will approach the particle pair within this
time, the pair will be surrounded by more particles, a cluster is formed and it will get increasingly difficult to release
the particles from the cluster. The cluster is thus growing and ideally there is complete phase separation
into a dense region inside the cluster and a dilute region outside. Clearly the travel time from a neighbouring
particle to the initial pair is governed by the particle motility $v_0$, hence the particle phase separation is
purely induced by motility and is therefore called
motility-induced phase separation (MIPS). Conversely, in equilibrium (i.e. without motility at $v_0=0$), purely repulsive
interactions will never lead to fluid-fluid phase separation.
Clearly MIPS is also favored by higher particle densities since then particle are closer anyway.
In a parameter space spanned by the particle density and the Peclet number there is either a one-phase or a two-phase region.
This is by now well-explored by simulation, theory and experiment and there are several reviews on this topic
\cite{Zoettl_Stark_JPCM,ourRMP,Speck_review,Tailleur_review,compte_rendue,Marchetti_review}.

\begin{figure}
  \centering
  \includegraphics[width=0.40\textwidth]{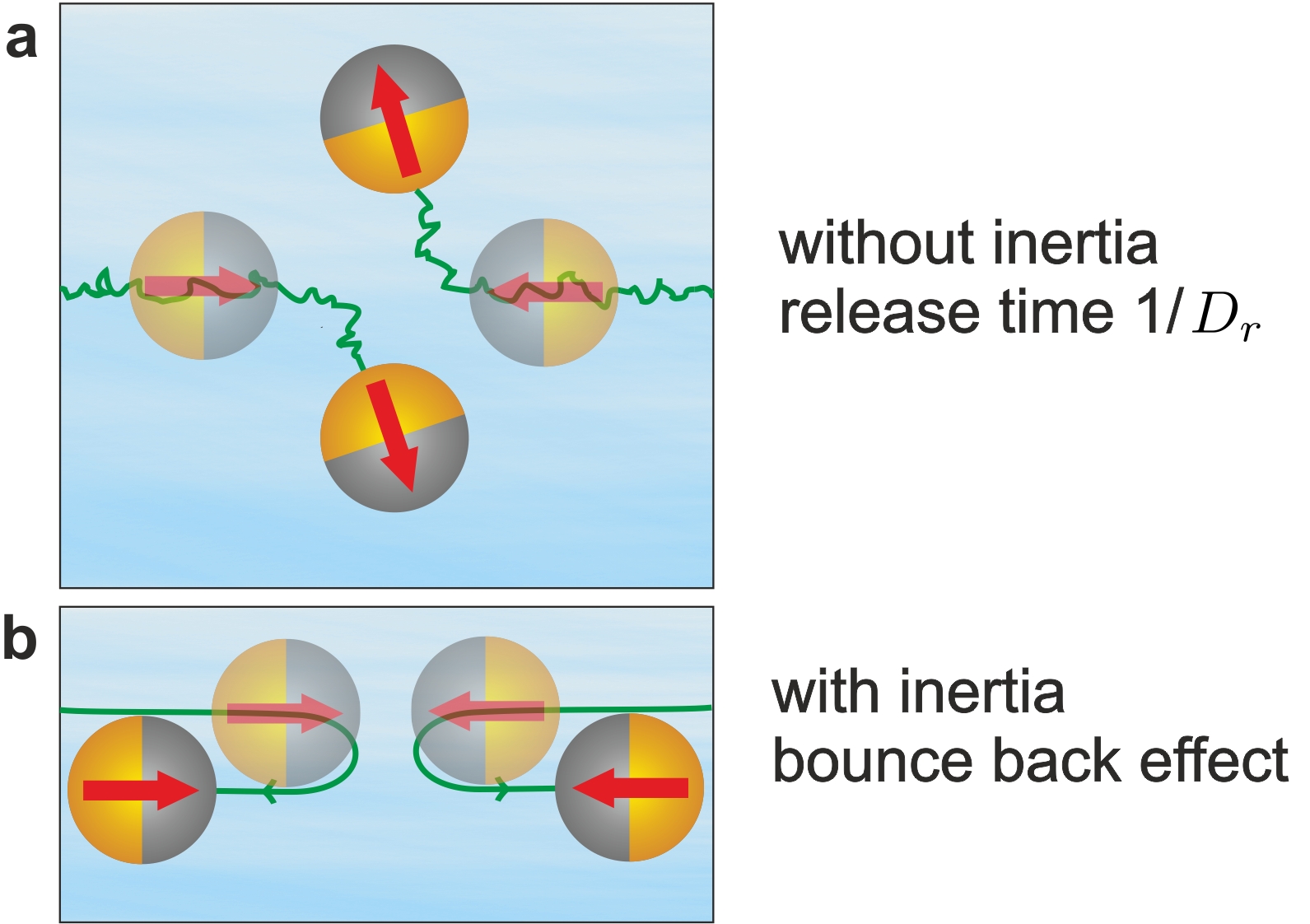}
  \caption{Sketch of a central elastic collision of two Janus particles with opposing self-propulsion velocities (red arrow):
a) active Brownian dynamics: the particles will stay almost touching for a typical time of $1/D_r$.
b) active Langevin dynamics: microflyers will bounce back such that they will not exhibit their terminal self-propulsion speed $v_0$. Initial positions are shown in light colors. The particle trajectories are rendered in green color.}
\label{Fig1}
\end{figure}

More recently MIPS has also seen for circle swimmers \cite{MIPS_circle1,MIPS_circle2}
and more complicated aligning interactions \cite{MIPS_aligning1,MIPS_aligning2}.
This shows that the occurence of MIPS itself is a very robust and general effect. Moreover the growth exponent of the cluster size
as a function of time
has been simulated \cite{Cates_PRL} and studied by theory \cite{Speck_Menzel_2014}. At the late stage,
for long times, a universal power law similar to the Cahn-Hilliard theory was found with a universal growth exponent of $1/3$.

\section{Active Langevin motion for self-propelled particles with inertia (``microflyers'')}

\subsection{Single particles}

We now generalize the equations of active Brownian to that of active Langevin motion including inertia both
for the translational and the rotational part, see
e.g.\  Refs.\ \cite{Baskaran2,Enculescu_Stark_PRL,Brady,Zippelius,Ldov,Shankar,Puglisi,Dauchot_Demery_PRL_2019,Das,Vuijk1,Vuijk2,um2019}, as
\begin{equation} \label{eq:24}
\begin{aligned}
m\ddot{\vec{r}} + \gamma(\dot{\vec r}) &= \gamma v_0 \hat n + \vec{f} (t)\\
J \ddot\phi + \gamma_R \dot\phi  &= M+g(t)
\end{aligned}
\end{equation}
Here, $m$ is the particle mass and $J$ the moment of inertia. In case the orientational relaxation is fast,
one may consider the limit of  vanishing moment of inertia,
$J=0$  \cite{Enculescu_Stark_PRL,Brady,Zippelius,Das}.
For $M=0$ we recover active Langevin motion for linear microflyers, for $M\not=0$ these are circle flyers.
Now the particle velocity and the orientation are not necessarily proportional.

On top of the parameters characterizing a Brownian circle swimmer discussed in \ref{CIIb},
there are now two more system parameters.
These can be best put in terms of two additional relaxation times scales due to finite moment of inertia and mass.
The {\it orientational relaxation time} upon which an  angular velocity relaxes due to the finite moment of inertia
is given by
\begin{equation} \label{eq:tau_r}
\tau_r=J/\gamma_R
\end{equation}
In case the orientational relaxation is fast, $\tau_r \to 0$,
one may consider the limit of  vanishing moment of inertia,
$J=0$ which was assumed in several recent studies  \cite{Enculescu_Stark_PRL,Brady,Zippelius,Das}.
Correspondigly there is also a second time, the {\it translational relaxation time},
\begin{equation} \label{eq:tau_r}
\tau=m/\gamma
\end{equation}
upon which the translational velocities relax due to the finite mass $m$.

With suitable basic length and time scales of the persistence random walk, $\ell_p$ and $\tau_p$,
we can state the four independent system parameters of Eq.\ (\ref{eq:24}) as three basic dimensionless {\it delay numbers\/}
${\cal D}_0=\tau_r/\tau_p$, ${\cal D}_1=\omega_s\tau_r$, ${\cal D}_2=\tau_r/\tau$ plus the Peclet number
$Pe$ defined in Eq.\ (\ref{eq:Peclet}) which contributes a time scale for the translational Brownian motion.

Some analytical solutions of the active Langevin motion model are given in Ref.\ \cite{Ldov} which we briefly review here.
There is an analytical result for the noise-averaged and time-resolved
displacements and mean-square-displacements \cite{Jahanshahi}. Four different regimes can be identified
where the MSD exhibits different power laws: ballistic for very short times, then diffusive due to solvent noise,
then ballistic again due to self-propulsion and then diffusive again for very long times due to
randomizing particle orientation. The long-time
translational self-diffusion coefficient $D_L$ can be calculated as

\begin{equation} \label{eq:Ldov8}
D_{\mathrm{L}} = D + \frac{{v_0}^2}{2}{\frak t}\left( {\tau _{\mathrm{r}},{\frak D}_0,{\frak D}_1} \right)
\end{equation}

This equations has a similar structure than the result for overdamped dynamics, see Eq.\ (\ref{eq:20b}),
insofar that is is a superposition of the translational diffusion and an active term proportional to $v_0^2$.
The corresponding time scale is given by

\begin{equation} \label{eq:Ldov9}
{\frak t}\left( {\tau _{\mathrm{r}},{\frak D}_0,{\frak D}_1} \right) = \tau _re^{{\frak D}_0}{\mathrm{Re}}\left[ {{\frak D}_0^{ - \left( {{\frak D}_0 - i{\frak D}_1} \right)}{\mathrm{\gamma }}\left( {{\frak D}_0 - i{\frak D}_1,{\frak D}_0} \right)} \right]
\end{equation}
where $Re$ denotes the real part and $\gamma(y,z)$ is the lower incomplete gamma function.
Interestingly, the time scale (\ref{eq:Ldov9}) does {\it not} depend on the translational relaxation time $\tau$ but it depends explicitly
on the rotational relaxation time $\tau_r$. This is in contrast to equilibrium ($v_0=0$) where $D_L$ neither depends
on  $\tau$ nor on $\tau_r$. In fact, the dependence of long-time diffusion on the moment of inertia is pretty strong as

\begin{equation} \label{eq:Ldov14}
D_{\mathrm{L}} = D + v_0^2\sqrt {\frac{\pi }{{8D_{\mathrm{r}}\xi _{\mathrm{r}}}}} \sqrt J + {\cal O}\left( {\sqrt {J^{ - 1}} } \right)
\end{equation}
As an application animals can hardly change mass but they can change their moment of inertia during motion. So this may provide a
strategy to sample space more quickly.

Moreover the zero-time velocity
correlation, i.e. the mean kinetic energy, can be calculated as

\begin{equation} \label{eq:Ldov5}
\overline{{\dot{\vec{r}}}^2} = 2D{\mathrm{/}}\tau + {\frak f}\left( {{\frak D}_0,{\frak D}_1,{\frak D}_2} \right)v_0^2
\end{equation}

with

\begin{align} \label{eq:Ldov7}
\begin{array}{c}{\frak f}\left( {{\frak D}_0,{\frak D}_1,{\frak D}_2} \right) = {\frak D}_2e^{{\frak D}_0}{\mathrm{Re}}\left[ {{\frak D}_0^{ - \left( {{\frak D}_0 - i{\frak D}_1 + {\frak D}_2} \right)}} \right.\\ \left. { \times {{\gamma }}\left( {{\frak D}_0 - i{\frak D}_1 + {\frak D}_2,{\frak D}_0} \right)} \right]\end{array}
\end{align}
The first term in (\ref{eq:Ldov5}) is the equilibrium solution for a passive particle ($v_0=0$) and the second arises from the active motion.
The latter is proportional to $v_0^2$, i.e. the kinetic energy injected by the propulsion.

Remarkably the orientational correlation function $C(t)$ is a {\it double exponential\/}
\begin{equation} \label{eq:20cinertia}
C(t) = \overline{\hat{n}(t)\cdot\hat{n}(0)} = \cos{\omega_s t}\; \exp{(-D_r(t-\tau_r(1-e^{-t/\tau_r})))}
\end{equation}
emphasizing again the important role of rotational inertial relaxation since the time scale $\tau_r$ enters here explicitly.

Finally one finds for the
delay function which measures how the dynamical changes of orientation and velocities are correlated 
\begin{align} \label{eq:20delay2}
\begin{array}{c}d(t) = c(t,0) - c(0,t) = \\ \overline{{{\dot{\vec{r}}}(t) \cdot {\hat n}}(0)} - \overline{{{\dot{\vec{r}}}(0) \cdot {\hat n}(t)}}  = v_0{\frak D}_2e^{{\frak D}_0}{\frak D}_0^{({\frak D}_2 - {\frak D}_0)}e^{ - t/\tau }\\ \times {\mathrm{Re}}\left[ {{\frak D}_0^{i{\frak D}_1}\left( {{\frak D}_0^{ - 2{\frak D}_2}{{\gamma }}\left( {{\frak D}_0 - i{\frak D}_1 + {\frak D}_2,{\frak D}_0} \right)} \right.} \right.\\ - e^{2t/\tau }{\frak D}_0^{ - 2{\frak D}_2}{{\gamma }}\left( {{\frak D}_0 - i{\frak D}_1 + {\frak D}_2,{\frak D}_0e^{ - t/\tau _{\mathrm{r}}}} \right)\\ - {{\gamma }}\left( {{\frak D}_0 - i{\frak D}_1 - {\frak D}_2,{\frak D}_0e^{ - t/\tau _{\mathrm{r}}}} \right)\\ \left. {\left. { + {{\gamma }}\left( {{\frak D}_0 - i{\frak D}_1 - {\frak D}_2,{\frak D}_0} \right)} \right)} \right].\end{array}
\end{align}
By definition, this function is zero for $t=0$ and is then strictly positive exhibiting a maximum after a typical
characteristic delay time. This shows that on average first the particle orientation will change and then
the particle velocity will follow on the scale of this delay time.

We close this section with two remarks:
First, for $J=0$ there is an equivalence to overdamped particle motion
in in a harmonic potential as can easily be seen by replacing the role of velocities and positions in the equations of motion.
This mapping has lead to some other exact results for the dynamics obtained by Malakar and coworkers  \cite{Kumar}.
Second, there are more complicated models to describe an additional alignment between particle orientation and velocity
which is ignored in Eq.\ (\ref{eq:24}). This seems to be a more realistic description of granular hoppers
\cite{Reference_[16]_in_Dauchot_Demery_PRL_2019,Dauchot_Demery_PRL_2019} but lacks an analytical solution.
It has been applied to
inertial active particles in a harmonic external potential recently \cite{Dauchot_Demery_PRL_2019}.
This extra term involves an additional torque such that the orientational dynamics in
Eq.\ (\ref{eq:24}) can be written for $J=0$ as
\begin{equation} \label{eq:Dauchot}
\gamma_R \dot {\hat{n}}  =  \zeta ( {\hat{n}} \times \dot{\vec r} ) \times {\hat{n}} +  ({\vec M}  + { \vec g}(t) ) \times {\hat{n}}
\end{equation}
where $\zeta$ is a coupling coefficient.

\subsection{Self-propulsion of microflyers in non-inertial frames}

The equations of motion Eq.\ (\ref{eq:24}) can be generalized to non-inertial frames
to describe self-propulsion on rotating disks or on oscillating plates, for example,
as has been discussed recently \cite{Loewen_noninertial}. The new phenomenon for active Langevin motion in this set-up is that
additional fictitious forces have to be added to the equations of motion
if the equations are expressed in the non-inertial frame.
We briefly illustrate this for a planar rotating disk and for an oscillating plate.

\subsubsection{Rotating disk}

On a planar disk
rotating around the $z$-axis with a constant angular velocity $\omega$, the equations of motion for
a particle self-propelling on the $xy$-plane read in the laboratory frame as

\begin{equation} \label{eq:26}
 \begin{aligned}
m\ddot{\vec{r}} + \gamma(\dot{\vec r}-\vec{\omega}\times \vec r) &= \gamma v_0 \hat n + \vec{f} (t)\\
J \ddot\phi + \gamma_R (\dot\phi - \omega) &= M+g(t)
 \end{aligned}
\end{equation}

The ingredient here is that friction is proportional to the {\it relative} velocity and the {\it relative} angular
velocity in the rotating non-inertial frame. If expressed as equations of motion in the rotating frame,
the friction term is getting easier but  additional centrifugal and Coriolis forces need to be included.

\begin{figure}
  \centering
  \includegraphics[width=0.48\textwidth]{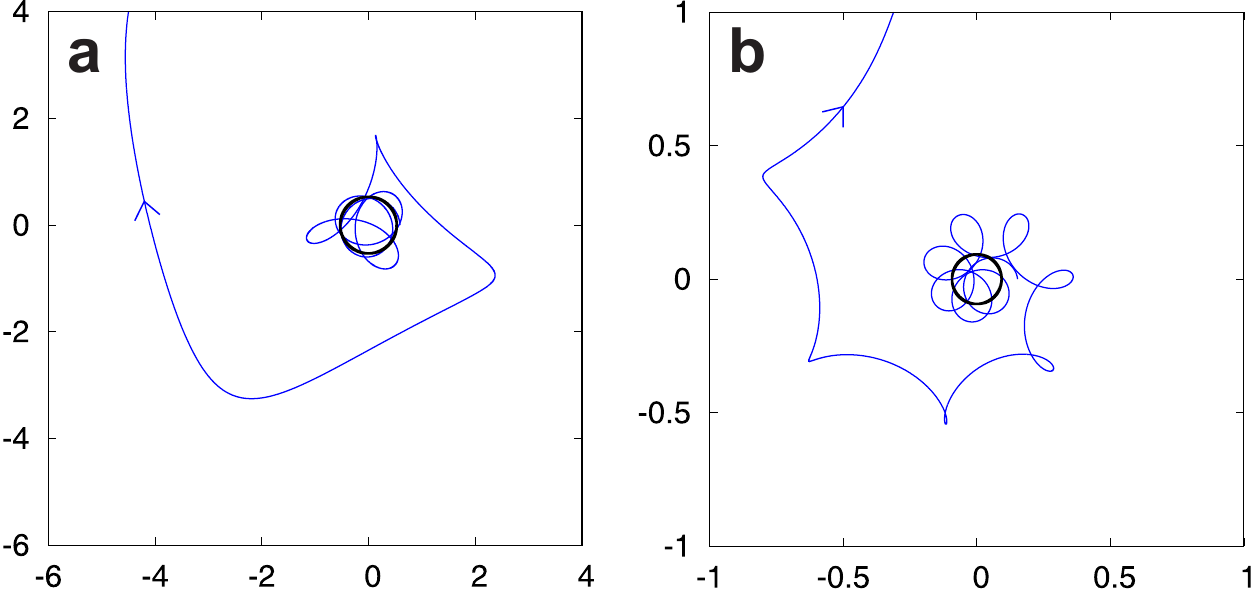}
  \caption{Typical trajectories from the analytical solution of the noise-free Eq. (\ref{eq:26}) in the laboratory frame.
The unstable rotation with radius $b$ is indicated as a black circle. The trajectory approaches a logarithmic spiral. 
The length unit is $v_0/\omega$ and
the parameters are: $\gamma / m\omega = 2.5$ a) $\omega_s/\omega = 1$ and b) $\omega_s/\omega = 4$}
\label{Fig2}
\end{figure}

In the noise-free case, the solution can be found analytically and is given by a superposition of three terms:
two logarithmic spirals which spiral inwards and outwards with different rates and a constant rotation around
the rotation origin
with radius

\begin{equation} \label{eq:30}
b = \frac{\gamma v_0}{\sqrt{m^2 (\omega+\omega_s)^4 + \gamma^2 \omega_s^2}} 
\end{equation}
The special circular solution can be understood as arising from a balance of the centrifugal force, the self-propulsion force
and the friction force in the rotating frame. However, it is unstable with respect to the out-spiraling part
which stems from the action of the centrifugal force. The actual trajectories looks pretty complex
in the laboratory frame. An example is shown in Figure \ref{Fig2}.
For an overdamped system, these fictitious forces do not exist and a particle does
not suffer from the centrifugal expulsion from the rotation center.

\subsubsection{Oscillating plate}

We now consider active Langevin motion on a two-dimensional  oscillating  plate.
The oscillating plate constitutes a linearly accelerated frame of reference as described
with a time dependent distance vector
\begin{equation} \label{eq:oscillating}
{\vec R}_0(t) = D_p \cos(\omega_p t) {\vec e}_x
\end{equation}
between the origins of the inertial laboratory frame and the non-inertial frame.
Here $D_p$ is an oscillation amplitude, $\omega_p$ is the oscillation frequency of the plate and the oscillation
is taken along the $x$-axis without loss of generality.
The equations of motion in the laboratory frame are

\begin{align} 
m \ddot{\vec{r}} + \gamma (\dot{\vec{r}} - \dot{\vec{R}}_0 (t)) &= \gamma v_0 \hat n + \vec{f} (t)\label{eq:L2}\\
\gamma_R \dot{\phi} &= M+ g(t)\label{eq:L3}
\end{align} 
and the solution in the noise-free case can be obtained analytically as a superposition of harmonic terms
with frequencies $\omega_p$ and $M/\gamma_R$ in $x$-direction. One term is exponentially damped with rate $\gamma/m$,
another is persistent and undamped. The effect of noise, the noise-averaged trajectory and the MSD
can be calculated analytically following the analysis proposed in \cite{Loewen_noninertial}. For large
$\omega_p$ (i.e. $\omega_p >> M/\gamma_R$) there is an enormous amplification of the oscillation amplitude
due to the fictitious centrifugal force.

\subsection{Collective effects: MIPS}

\begin{figure*}
  \centering
  \includegraphics[width=1.0\textwidth]{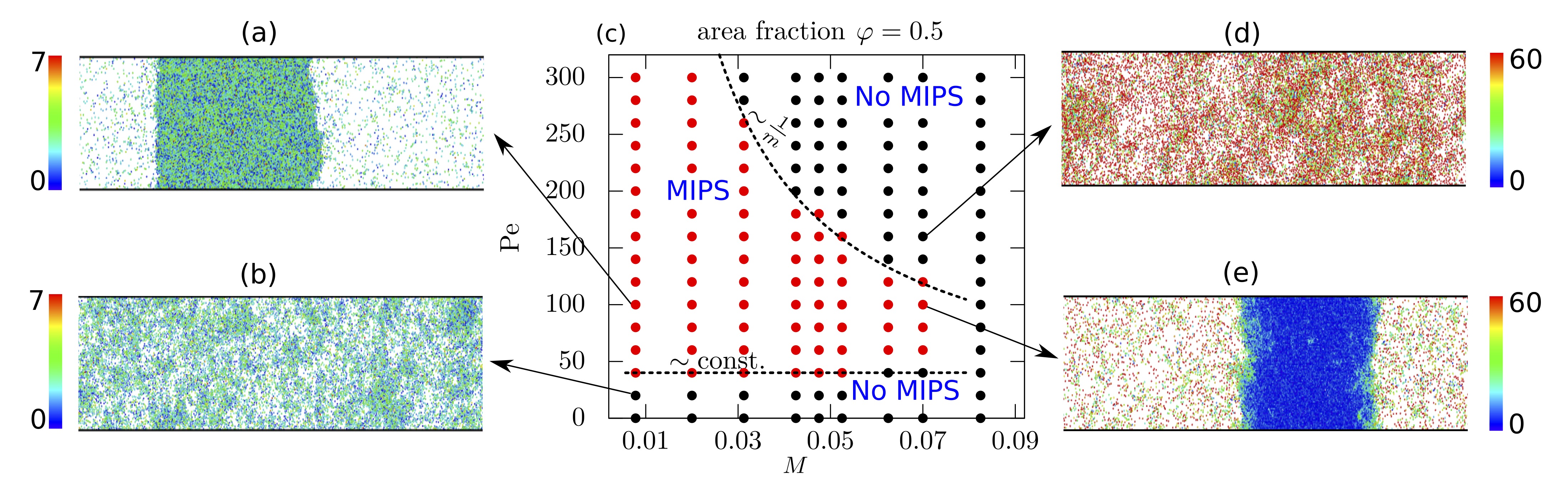}
  \caption{Nonequilibrium phase diagram at an area fraction of $\phi= 0.5$ in the plane spanned by the particle mass and the self-propulsion speed (arbitrary units) (c). Panels (a), (b), (d), and (e) represent simulation snapshots in slab geometry at state points indicated in the phase diagram. Colors represent kinetic energies of individual particles in units of $k_B T$. A hot-cold coexistence is visible in panel (e). Dashed lines in (c) show scaling predictions for the phase boundary between the homogeneous and phase-separated state. From Ref.\ \cite{Suvendu_Liebchen}.}
\label{Fig3}
\end{figure*}

Since motility-induced phase separation (MIPS) is in general a pretty robust effect,
it is expected to occur also for inertial active particles provided the inertial effects are not too large.
In fact, recent studies \cite{Suma1,Suvendu_Liebchen} have explored the active Langevin model in this regard and found that
inertia as modelled by the many-body generalization of Eq.\ (\ref{eq:24})
is unfavorable for MIPS. An example is shown in Figure \ref{Fig3} where the phase separation separatrix
is shown for fixed density in the parameter space spanned by the Peclet number $Pe$ and the particle mass $m$. In this case,
the moment of inertia $J$ and the external torque $M$ were both set to zero.
In fact, beyond a critical mass, there is no phase separation at all. Generally speaking
this has to do with the additional fluctuations which occur in the active system introduced by inertia.
If we take the intuitive picture for MIPS in underdamped systems described in \ref{CIIc}, it is now getting different: with inertia,
a centrally colliding particle pair will {\it bounce back} rather than staying static and Brownian
as it does in the overdamped case, see Figure \ref{Fig1}b. This will destroy the nucleus for subsequent particle
aggregation more than it does in the overdamped case and therefore MIPS is unfavored by inertia.
Moreover there is a re-entrant one-phase region if the activity (or Peclet number) is increased
which is strongly amplified by inertia.

\begin{figure}
  \centering
  \includegraphics[width=0.48\textwidth]{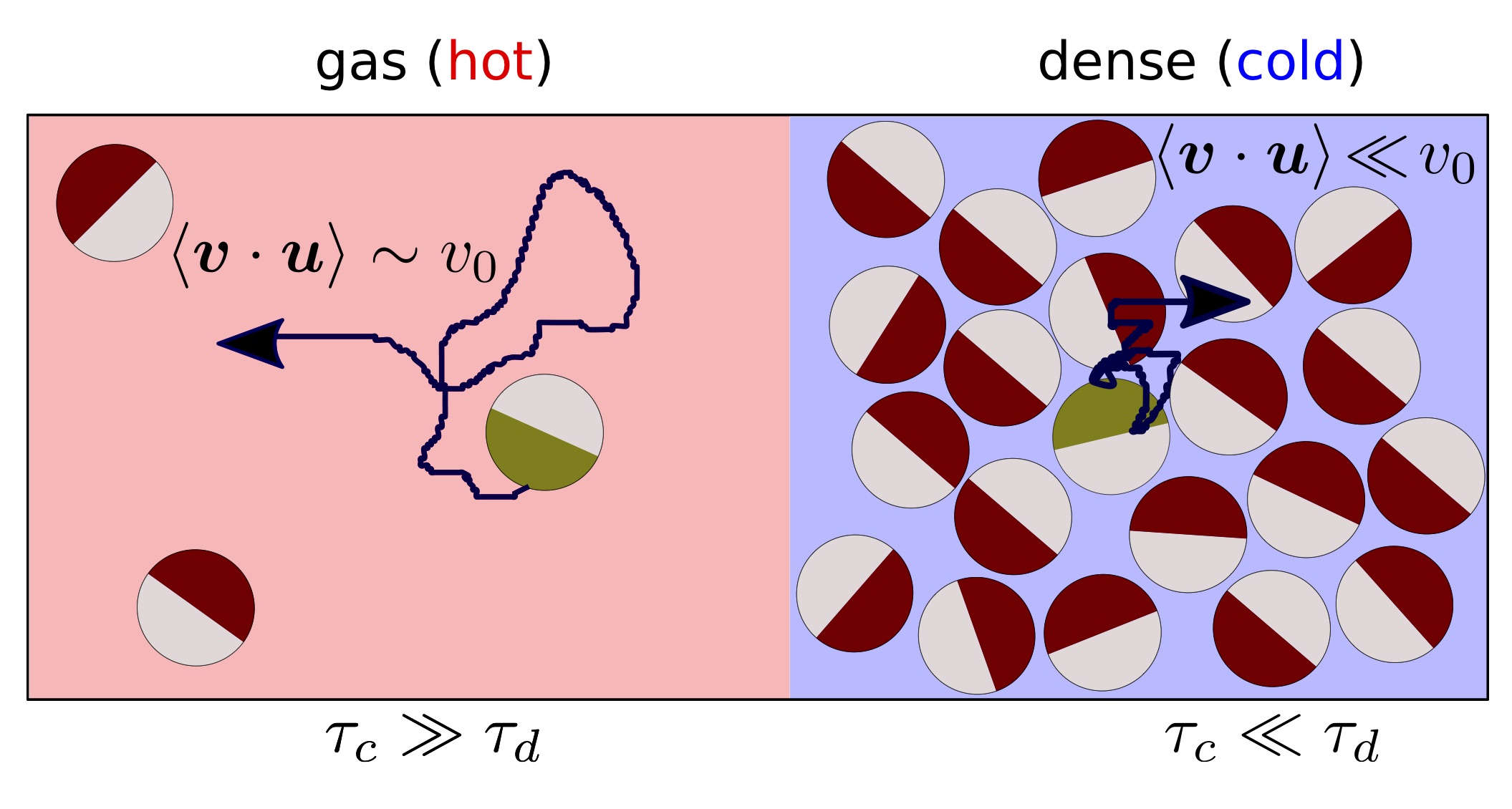}
  \caption{Scheme of the phase-separated state associated with a hot-cold coexistence in underdamped active particles. Particles self-propel with the colored cap ahead (brown; greenish for the tagged particle). Active particles move with $v_0$ in the gas phase, but can be an order of magnitude slower in the dense phase. From Ref.\ \cite{Suvendu_Liebchen}.}
\label{Fig4}
\end{figure}

But when MIPS occurs for inertial active particles there is novel effect that does not occur for overdamped systems:
the two coexisting phases exhibit a different temperature \cite{Suma2,Suvendu_Liebchen}. Here temperature is
defined via the mean kinetic energy of the particles. Figure \ref{Fig4} shows the underlying principle.
In contrast to granulates where a similar effect has been found \cite {Ref_[33]_in_Suvendu_Liebchen},
particle collisions are elastic here but the self-propulsion makes a collision between two particles like two bouncing balls
hitting eachother, see again Figure \ref{Fig1}b. Thus particles will never possess a velocity along their orientation
when there are many subsequent inter-particle collisions. Therefore the dense region is "cool" (in terms of kinetic temperature)
while in the dilute region particles will accelerate until they have almost reached their terminal velocity $v_0$.
Hence the dilute region is "hot". The temperature difference between the dilute and dense regions
can be huge up to a factor of 100. Note that there is no flux of heat at the fluid-fluid interface
but a stable thermal gradient will be established there.

Finally the growth of particle clusters during the phase separation process has been explored by computer simulation
\cite{Suvendu_Liebchen}.
These calculations reveal that the cluster growth exponent ist significantly lower than the universal
value of $1/3$ found in the overdamped case \cite{Cates_PRL,Speck_Menzel_2014} proving that inertia can
qualitatively change the physics of the collective phenomena.

In summary, in the active Langevin model, there are three basic effects which are caused by inertia as far as motility-induced
phase separation is concerned: the phase separation is shifted towards higher Peclet number and is finally destroyed completely
as the particle mass is increased. Second, the kinetic temperature is different in the two coexisting phases in stark
contrast to equilibrium thermodynamics where phase coexistence implies equality of temperatures. Third, the cluster
growth exponents is smaller than the universal value $1/3$ valid for overdamped systems.

\section{Summary of translational and orientational dynamical autocorrelation functions}\label{CIV}



In Table I we summarize the different cases discussed so far in terms of the
translational and orientational dynamical autocorrelation functions $Z(t)$ and $C(t)$. We distinguish
between a simple single exponential decay with one decay time and more complicated behavior such as an oscillatory decay
(valid for circle swimmers) or double exponentials (valid for microflyers with $M=0$ and $J>0$).
The passive cases are listed as references. For passive Langevin dynamics
\begin{equation} \label{eq:24iii}
m\ddot{\vec{r}}(t) +  \gamma  \dot{\vec r}(t) =  \vec{f} (t)
\end{equation}
with $\vec{f} (t)$ denoting white Gaussian noise,
the  velocity autocorrelation function $Z(t)$ decays as a simple exponential with a decay time
$m/\gamma$ and the orientational dynamics is decoupled from this equation. For $J=0$ and $M=0$,
the orientational correlation is single exponential, but not for $J>0$ where it is a double exponential
or for $M>0$ where it is oscillatory.

\begin{table*}[]
\begin{tabular}{|c|c|c|c|c|}
\hline
                                                   & \multicolumn{2}{c|}{translational velocity correlation $Z(t)$} & \multicolumn{2}{c|}{orientational correlation $C(t)$} \\  \hhline{|=|=|=|=|=|}
\cellcolor[HTML]{FD6864}single particle            & single exponential      & \cellcolor[HTML]{DDDDDD}    more complicated     & single exponential      & \cellcolor[HTML]{DDDDDD}   more complicated     \\ \hhline{|=|=|=|=|=|}
passive Brownian motion                             & X                       & \cellcolor[HTML]{DDDDDD}                       & X                       &   \cellcolor[HTML]{DDDDDD}                     \\ \hline
active Brownian motion                                & X                       &  \cellcolor[HTML]{DDDDDD}                      & X                       &     \cellcolor[HTML]{DDDDDD}                   \\ \hline
Brownian circle swimmer                             &                         & \cellcolor[HTML]{DDDDDD}    X                  &                         &  \cellcolor[HTML]{DDDDDD}   X                    \\ \hline
passive Langevin $(J=M=0)$                       & X                       &         \cellcolor[HTML]{DDDDDD}                 & X                       &   \cellcolor[HTML]{DDDDDD}                  \\ \hline
passive Langevin $(J>0$ or $M>0)$            & X                       &        \cellcolor[HTML]{DDDDDD}                   &                         &  \cellcolor[HTML]{DDDDDD}   X       \\ \hline
active Langevin $(M=J=0)$                                       &                       &   \cellcolor[HTML]{DDDDDD}  X         &    X                     & \cellcolor[HTML]{DDDDDD}                 \\ \hline
active Langevin $(J\neq0)$                          &                         &  \cellcolor[HTML]{DDDDDD}   X                   &                         &  \cellcolor[HTML]{DDDDDD}   X      \\   \hhline{|=|=|=|=|=|}
\cellcolor[HTML]{FD6864}many interacting particles  &                         &  \cellcolor[HTML]{DDDDDD}         &                         &       \cellcolor[HTML]{DDDDDD}             \\ \hhline{|=|=|=|=|=|}
non-aligning $(M=0)$                                   &                         & \cellcolor[HTML]{DDDDDD}    X                  & X                       &      \cellcolor[HTML]{DDDDDD}                  \\ \hline
aligning interaction                                       &                         &         \cellcolor[HTML]{DDDDDD}     X             &                         &  \cellcolor[HTML]{DDDDDD}   X               \\ \hline
\end{tabular}
\caption{Summary of the behavior of translational and orientational correlation functions $Z(t)$ and $C(t)$
for different situations of single and many passive and active particles. In particular a simple
single exponential decay is indicated. All four different combinations do occur,
the models belonging to the corresponding classes are listed.}
\end{table*}

For many particles with non-aligning interactions (at vanishing external torque, $M=0$),
the orientational correlation function is still a single exponential, but the translational correlation function is highly nontrivial
(even for passive particles \cite{Hansen_Mac_Donald_book_Theory_of_Simple_Liquids}) while for aligning interactions also the orientational dynamics is complicated
\cite{JanDhontbook_Dynamics_of_colloids}.

\section{Experimental realization}\label{CV}

\subsection{General}\label{CV1}

Inertial effects are getting relevant in particular for two different situations:
i)  for {\it macroscopic\/} self-propelled particles,
ii) for mesoscopic particles on the colloidal scale moving
in a medium of {\it low viscosity} (such as a gas).

Regarding the first situation, one of the
best realization of our model equations (\ref{eq:24}) for active Langevin dynamics can be found for active granulates
\cite{Ramaswamy,Dauchot_Frey_PRL_2013,Poeschel,Deblais,Tsimring,Chate,Patterson,Dauchot2,Mayya1,Mayya2}.
Typically these are hoppers with
a Janus-like body or with tilted legs. In order to achieve self-propulsion, these macroscopic
bodies are either placed on a vibrating table or are equipped with an internal vibration motor
("hexbugs") \cite{Dauchot_Demery_PRL_2019}. It has been shown that
the dynamics of these hoppers is  well described by
active Brownian motion with inertia \cite{Walsh,Dauchot_Brownian,Ldov}. Since they are macroscopic, inertia is relevant.
The fluctuations can be fitted to Brownian forces and imperfections in the particle symmetry
will make them circling ($M\not= 0$). Therefore, they are ideal realizations of our model equations (\ref{eq:24}) but there is a caveat for certain granulates insofar as the
additional aligning torque described in Eq.\ (\ref{eq:Dauchot}) needs to be included.
Moreover, there is no major difficulty in placing granulates on a turntable or on an oscillating
plate so that effects arising from a non-inertial frame are directly accessible.
There is a plethora of other examples of macroscopic self-propelling objects which are dominated by inertia.
These include mini-robots \cite{Leyman,Mijalkov},  flying whirling fruits \cite{Ldov}
as well as cars, boats, airplanes, swimming and flying animals \cite{beetles=Ref[25]_in_Mandal}
and moving pedestrians, bicyclists and vehicles \cite{pedestrians,zhang2013,chowdhury2000}.

Regarding the second situation, dust particles
in plasmas ("complex plasmas")
can be made active \cite{Bartnick}. They exhibit underdamped dynamics due to the presence of the neutral gas
\cite{Ivlev_Morfill} and are therefore highly inertial. Another example in nature are fairyflies which belong
to the smallest flying insects on our globe and have a size of several hundreds of microns.


{\subsection{Vibrated granulates in particular}}\label{CV2}

\begin{figure}
  \centering
  \includegraphics[width=0.48\textwidth]{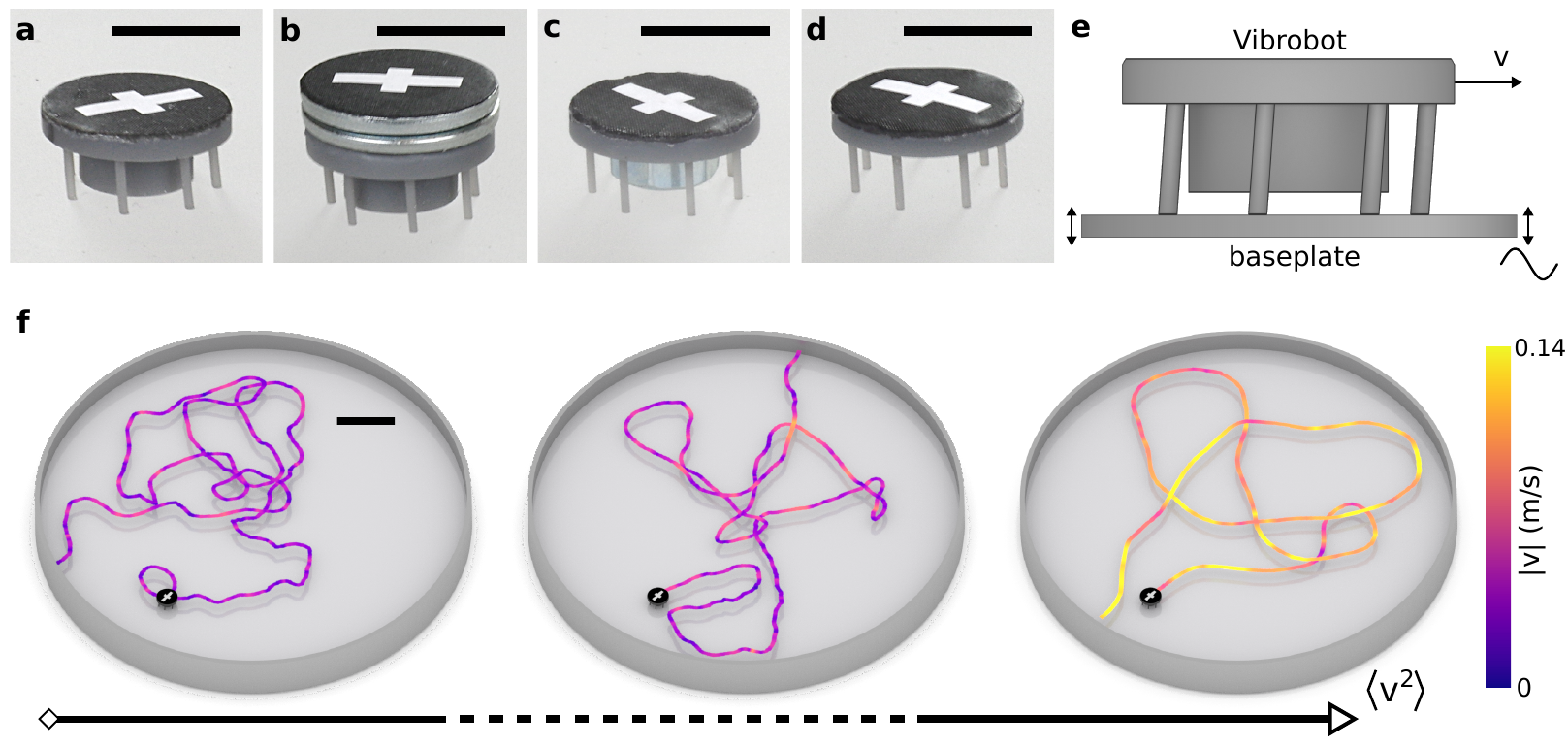}
  \caption{ \textbf{3D printed particles, setup and trajectories.} \textbf{a} \textit{Generic particle} \textbf{b} \textit{Carrier particle} with an additional outer mass. \textbf{c} \textit{Tug particle} with an additional central mass. \textbf{d} \textit{Ring particle} without a central core. \textbf{e} Illustration of the mechanism with a generic particle on a vibrating plate. \textbf{f} Three exemplary trajectories with increasing average particle velocities. Particle images mark the starting point of each trajectory. The trajectory colour indicates the magnitude of the velocity. From Ref.\ \cite{Ldov}.}
\label{FigV21}
\end{figure}

As already mentioned, a direct realization of active Langevin motion is obtained for vibrated granulates. Interestingly, particles
can be prepared with different mass and different moments of inertia, see Ref.~\cite{Ldov} which we describe now in more detail. Exposed to a vibrating plate,
they perform self-propulsion in two dimensions, see Figure \ref{FigV21} for the experimental realization
and typical particle trajectories. The actual particle velocity along the trajectories is not constant
but fluctuates and the mean-square displacements and orientational correlations are in good agreement with the active Langevin model
when some parameters are fitted to the experiments. In particular, the velocity distribution function
has a peak around the self-propulsion velocity.

\begin{figure*}
  \centering
  \includegraphics[width=1.0\textwidth]{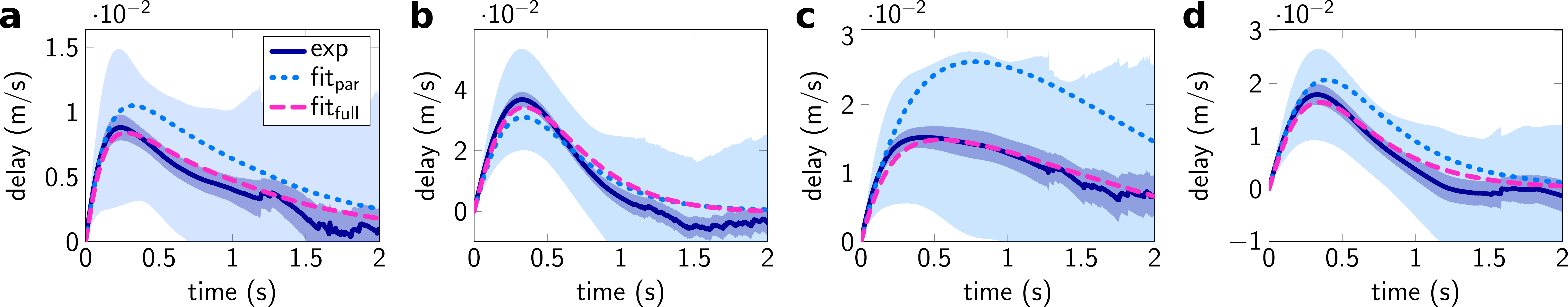}
  \caption{Delay functions $d(t)$ for the \textbf{a} generic, \textbf{b} carrier, \textbf{c} tug and \textbf{d} ring particle. The solid dark blue shows experimental data, the dashed magenta and dotted light blue curve plot theoretical results, with a different way of fitting. Experimental uncertainties are expressed as the standard deviation in light or dark blue regions. From Ref.\ \cite{Ldov}.}
\label{FigV22}
\end{figure*}

Interestingly experimental data for the inertial delay as embodied in the delay correlation function
$d(t)$ (see Eq.\ (\ref{eq:20delay2})) are both in qualitative and quantitative agreement with the theoretical result.
An example is shown in Figure \ref{FigV22}.
Indeed the theoretical prediction is confirmed by the experimental data averaged over the noise. This documents that first the
direction of self-propulsion is changed and the velocity follows. As this delay effect is missing in the overdamped case, it is
caused completely by inertia. It is exactly this inertial delay effect which is used by oversteering  racing cars to get around corners. We finally remark that, as compared to hexbugs, no self-aligning forces  \cite{Dauchot_Demery_PRL_2019} need to be incorporated to get a reasonable fit of the data.\\

\section{Perspectives}\label{CVI}

In the flourishing field of active matter most of the previous
investigations used simple overdamped dynamics such as active Brownian motion
to model microswimmers and meso-scale self-propelled colloids. If it comes to more macroscopic
active particles (granulates or minirobots) or to motion in a gas ("microflyers"), inertial effects become relevant.
Therefore it is expected that future research will
include more and more aspects of inertia also on a more fundamental level. In this article we have mainly
touched the basic model description of inertial active matter and their realization in granulate experiments.

Future activities and promising perspectives are expected along the following directions:

First, granulate particles will play a leading role as {\it paradigmatic realizations\/} for active matter models.
Since they are macroscopic, the particles can more easily be manipulated and changed. The particle shape can be easily changed by
macroscopic 3d-printing and different particle interactions can be established. In detail:

i) The whole field of {\it charged active matter} which unifies strongly coupled unscreened Coulomb systems \cite{Levin_review}
and active matter can be realized by charging granulates,
either by tribo-electric effects and by preparing macroscopic charged particles \cite{Haeberle}. This is possibly a
better controlled charged system than that of charged active dusty plasmas which require nonequilibrium ionic fluxes.

ii) Dipolar active particles with permanent magnetic dipole moments are not easy to synthesize on the colloidal level
\cite{Casic,Erbe,Nourhani,VandeWalle} but can directly
be realized by equipping granulates with little permanent magnets.

iii) We are just at the beginning to study active polymers as colloidal chains. While there is an increasing number
of simulation and theoretical studies, see \cite{Winkler_Review} for a recent review, experimental studies with active polymers and
colloidal chains in an active bath are sparse \cite{Yan_2016_Ref[38]_in_EPL_2018}. The effect of
inertial dynamics on active polymers needs to be understood better and future experiments and simulations are expected in this direction.

iv) It would be highly interesting to study {\it active surfactants}, to couple the field of surfactants with active matter.
Chain-like vibrated granulates with a head and tail part composed of rotators with a different rotation sense
\cite{Ref_[10]_in_Ldov} can be prepared and brought into motion to explore the
dynamics at a surfactant interface.

v) Studying granulates on a vibrating
{\it structured substrate\/} will induce anisotropic active motion which has not yet been studied by theory either.

vi) Time-dependent propulsion strengths when the propulsion velocity depends explicitly on time can be realized by granulates,
for example by modulating the shaking amplitude on demand.
For overdamped systems, some analytical results were obtained for time-dependent propulsions \cite{Babel_1}
but inertia is expected to induce new lag-effects in the dynamics.

Second, the motion of inertial active particles will be explored in various {\it confining geometries\/}
(harmonic trap, confining walls).  It has already been shown that in a simple harmonic confinement
there are novel dynamical effects \cite{Dauchot_Demery_PRL_2019}. This will be even more complex for more complicated confinements.

Third, {\it collective effects\/} of active Langevin dynamics needs to be explored more. More studies on MIPS including a non-vanishing moment of inertia and an external torque will be performed. The role of aligning interactions
needs to be understood better for inertial systems \cite{EPL_Ignacio,Dijkstra_aligning}.
Active crystallization will be studied where inertia
provides a latent heat upon crystallization. Here we do not only need benchmarking experiments but also fundamental theory
including the Langevin dynamics. First attempts in terms of theory
have been done by generalizing the swim pressure to the inertial case \cite{Brady}
and to consider inertial terms in hydrodynamic approaches \cite{Puglisi,Ramaswamy_new} but certainly also microscopic
approaches such as mode-coupling theory \cite{MCT1,MCT2}  and dynamical density
functional theory \cite{DDFT} need to be extended to include inertia.
Next, active particles with inertia may provide little heat engines with a better efficieny than
their overdamped counterparts as energy is not damped away by the dynamics. We are just at the beginning to understand the principles
of entropy production \cite{FodorE,SeifertE}  and heat conversion \cite{SoodH,FodorH,SeifertH} in these systems.

Finally, inertia introduces some kind of {\it memory\/} to the particle motion, both for translational and orientational motion,
on the time scale of the inertial relaxation times $\tau$ and $\tau_r$. This is the prime
reason for delay effects relative to
overdamped active Brownian particles. There is a need to classify memory effects in general and to study whether or not
the behavior is similar to that in other system governed by memory \cite{Schilling_2019}.
One other example where memory effects are crucial is
an  active particle in a viscoelastic (non-Newtonian) fluid  such as a polymer solution \cite{Krueger}
or a nematic liquid crystal \cite{Wensink_Toner,Abdallah} where a notable increase of the rotational
diffusion coefficient has been found \cite{Ruben}. Another example is a
sensorial delay in the perception of artificial minirobot systems \cite{Leyman,Mijalkov,Ihle} which was shown to have  a
significant effect  on the clustering and swarming properties.\\

\section{Conclusions}\label{CVII}
In conclusion we have upgraded the standard model of active Brownian motion by including inertia in
both translational and orientational motion leading to the basic model of active Langevin motion.
We discussed single particle properties by the orientational and translational correlation functions
presenting some analytical results for this model. When comparing the model to experiments on vibrated granulates,
good agreement was found. We summarized some effects induced by inertia including an inertial delay
between self-propulsion direction and particle velocity, the tuning of the long-time self-diffusion by
the moment of inertia, the effect of fictitious forces in non-inertial frames, and the influence of inertia on
motility-induced phase separation. Since inertial effects will necessarily become relevant
for length scales between macroscopic and mesoscopic both for artificial self-propelled objects and for living creatures,
a booming future of inertial active systems is lying ahead.

\acknowledgments
I thank
Soudeh Jahanshahi, Christian Scholz,
Thomas Franosch, Thomas Voigtmann, Alexander Sprenger, Alexei V. Ivlev,
Frederik Hauke and Suvendu Mandal for helpful discussions and
gratefully acknowledge support by the Deutsche
Forschungsgemeinschaft (DFG) through grant
LO 418/23-1.

\bibliography{bibfileJCPpers}

\end{document}